\documentclass[aps,pra,twocolumn,superscriptaddress,longbibliography]{revtex4-1}
\usepackage{graphicx}
\usepackage{color}
\usepackage{amssymb}
\usepackage{amsmath}
\usepackage{bm}
\usepackage{mathrsfs}
\usepackage{float}
\usepackage[separate-uncertainty=true,multi-part-units=single]{siunitx}
\usepackage{epstopdf}
\usepackage{multirow}

\definecolor{mygreen}{rgb}{0,0.5,0} 
\definecolor{mygrey}{rgb}{0.5,0.5,0.5} 
\definecolor{myred}{rgb}{0.75,0,0} 
\definecolor{myblue}{rgb}{0,0,0.75} 
\definecolor{mymagenta}{cmyk}{0,1,0,0.12} 
\definecolor{mycyan}{cmyk}{1,0,0,0.12} 
\definecolor{myorange}{rgb}{1,0.5,0}

\newcommand{\E}{{\cal E}}

\newcommand{\be}{\begin{equation}}
\newcommand{\ee}{\end{equation}}
\newcommand{\bea}{\begin{eqnarray}}
\newcommand{\eea}{\end{eqnarray}}

\usepackage{hyperref}

\begin{document}

\title{
Strong light shifts from near-resonant and polychromatic fields: comparison of Floquet theory and experiment
}

\newcommand{\ICFOAddress}{ICFO-Institut de Ciencies Fotoniques, The Barcelona Institute of Science and Technology, 08860 Castelldefels, Barcelona, Spain}
\newcommand{\ICREAAddress}{ICREA -- Instituci\'{o} Catalana de Re{c}erca i Estudis Avan\c{c}ats, 08015 Barcelona, Spain}
\newcommand{\LENSAddress}{LENS-European Laboratory for Nonlinear Spectroscopy, Universit\`a di Firenze, 50019 Sesto Fiorentino, Italy}
\newcommand{\NataliAddress}{Lone Star College, University Park, Chemistry and Physics Department, Houston, TX 77070}
\newcommand{\ThomasAddress}{Koheron, Centre scientifique d'Orsay, Batiment 503, 91400 Orsay, France}

\author{Simon Coop}
\email{simon.coop@icfo.eu}
\affiliation{\ICFOAddress}
\affiliation{\LENSAddress}
\author{Silvana Palacios}
\affiliation{\ICFOAddress}
\author{Pau Gomez}
\affiliation{\ICFOAddress}
\author{Y. Natali Martinez de Escobar}
\affiliation{\NataliAddress}
\author{Thomas Vanderbruggen}
\affiliation{\ThomasAddress}
\author{Morgan W. Mitchell}
\affiliation{\ICFOAddress}
\affiliation{\ICREAAddress}

\date{\today}

\begin{abstract}
We present a non-perturbative numerical technique for calculating strong light shifts in atoms under the influence of multiple optical fields with arbitrary polarization. We confirm our technique experimentally by performing spectroscopy of a cloud of cold $^{87}$Rb atoms subjected to $\sim$ kW/cm$^2$ intensities of light at 1560.492 nm simultaneous with 1529.269 nm or 1529.282 nm.  In these conditions the excited state resonances at 1529.26 nm and 1529.36 nm induce strong level mixing and the shifts are highly nonlinear.  By absorption spectroscopy, we observe that the induced shifts of the 5P3/2 hyperfine Zeeman sublevels agree well with our theoretical predictions.. We propose the application of our theory and experiment to accurate measurements of excited-state electric-dipole matrix elements.
\end{abstract}

\maketitle

\section{Introduction}
Light shifts, or ac Stark shifts, are ubiquitous in optical trapping of atoms \cite{Grimm2000}. They can be exploited to determine atomic properties for fundamental physics \cite{Arora2007, Herold2012}, in sensing applications such as optical magnetometry they can be detrimental \cite{Novikova2001, Budker2007} or beneficial \cite{Jimenez2014}, they can be used to characterize optical traps \cite{Brantut2008}, and recently have been exploited for fine control and addressing of individual qubits in a trapped-ion quantum information processor \cite{Lee2016}. Light shifts due to both blackbody radiation and probe light are a limiting factor in the accuracy of modern optical atomic clocks \cite{Zanon2016, Ludlow2015}.

Light shifts are often calculated using second-order perturbation theory \cite{Lee2016, BransdenJoachain, Kaplan2002, Schmidt2016, Leonard2015, Costanzo2016, LeKien2013, Levi2016}, however this is not adequate in situations with strong nonlinear light shifts and non-negligible mixing of different hyperfine energy levels. Here we describe a non-perturbative semiclassical theory for calculating light shifts based on Floquet theory. The theory can accurately describe light shifts in a regime analogous to the magnetic Paschen-Back regime, i.e. a regime where the light shifts are large and there is strong mixing of the hyperfine levels. It can describe light shifts due to multiple lasers of arbitrary polarization with wavelengths close to atomic resonances, with the limitation that the different wavelengths must be related by a rational fraction. At the same time, the mathematics is considerably simpler than in perturbative treatments \cite{LeKien2013,Levi2016} and handles strong level mixing in a natural way, thus extending the possibilities of light-shift engineering, e.g. for state preparation \cite{Griffin2006}.

We test our theory by performing spectroscopy on the light-shifted $|5P_{3/2},  F, M \rangle$ magnetic sublevels in optically trapped $^{87}$Rb. Our experiment can resolve light shifts of individual magnetic sublevels, and we find that our theory correctly predicts the positions of all levels after calibration of the in-situ light intensity and polarization. We use a simple model of atoms in a dipole trap to explain the observed spectrum. The spectrum is sensitive to both the trapping light intensity and polarization and can be used for calibration of both.

The theory presented here has a potential application to measuring excited-state electric-dipole matrix elements. Precise knowledge of dipole matrix elements is important for e.g. optical clocks, testing atomic structure calculations \cite{Safronova2011}, and atomic parity non-conservation measurements \cite{Fortson1984, Noecker1988}. While the idea of using light shifts to measure dipole matrix elements is not new \cite{Sahoo2009}, our theory enables the possibility of quantitative comparison between theory and experiment in a regime of strong light shifts.

\section{Floquet theory of light shifts}
\label{section:theory}
Floquet's theorem states that the Schr\"{o}dinger equation
\begin{equation}
i \hbar \frac{\partial}{\partial t} \psi(t) = H(t) \psi(t)
\label{eq:schrodinger}
\end{equation}
with time-periodic Hamiltonian $H(t) = H(t+T)$ has solutions of the form $\psi(t) = \phi(t) e^{-i \omega t}$, 
where $\phi(t) = \phi(t+T)$ has the same periodicity as $H(t)$.  In the case of an atom in an oscillating external field, we have $H(t) = H_0 + V(t)$, where $H_0$ is the free-atom Hamiltonian and $V(t) = V(t+T)$ is a periodic potential.  $\psi(t)$ describes a dressed state of the Hamiltonian, with dressed energy $\hbar \omega$.  

To find the dressed states, it suffices to consider $\mathscr{U}(T,0)$, the time-evolution operator for one period of the potential, for which $\mathscr{U}(t+T,t) \psi(t) = \psi(t) e^{-i \omega T}$. The eigenstates of this operator are thus the dressed states $\psi_i(t)$, with eigenvalues $\exp[i \omega_i T]$. This determines $\omega_i$ up to additive multiples of $2 \pi /T$. When $T$ is several optical periods, $2 \pi /T$ is large relative to fine- and hyperfine-structure splittings, and $\hbar \omega_i$ can be unambiguously assigned by comparison against the bare energies. 

 To compute $\mathscr{U}(T,0)$, we use a numerical Euler method. First we partition $\mathscr{U}(T,0)$ into $N$ subintervals 
\begin{equation}
\mathscr{U}(t_N,t_0) = \mathscr{U}(t_N,t_{N-1}) ...  \mathscr{U}(t_2,t_1)\mathscr{U}(t_1,t_0).
\label{eq:product}
\end{equation}
then approximate $ \mathscr{U}(t_1,t_0) \approx  e^{-{i} H(t_0)(t_1 - t_0)/{\hbar}}$
to find
\begin{equation}
\mathscr{U}(T,0) \approx \prod_{n=0}^{N-1} e^{- {i} H(t_n) {T}/(N {\hbar})}
\label{eq:U}
\end{equation} 
where $t_n = n T / N$, and the order of the product must be as in Eq. \ref{eq:product}. We compute the above with increasing $N$ until convergence. 

Now we calculate the two terms in the Hamiltonian. We work in the basis $|nJFM\rangle$, in which the free-atom Hamiltonian $H_0$ is diagonal, with different $M$ states degenerate and 
\begin{equation}
\begin{aligned}
\langle nJF | H_0 | nJF \rangle = \langle nJ | H_0 | nJ \rangle + \frac{1}{2} \hbar A_{nJ} K \\ 
+ \; \hbar B_{nJ} \frac{ {3}K(K+1)/{2}  - 2I(I+1)J(J+1) }{2I(2I -1)2J(2J-1)}
\end{aligned}
\end{equation}
where $K \equiv F(F+1) - I(I+1) - J(J+1)$ and the hyperfine constants $A_{nJ}$ and $B_{nJ}$ for $^{87}$Rb are taken from  \cite{Safronova2011}. Fine-structure energies $\langle nJ | H_0 | nJ \rangle$ are taken from the NIST atomic spectra database \cite{NIST_ASD}.

We describe the interaction between the atoms and the light in the electric-dipole approximation, so
\begin{equation}
V(t) = -\mathbf{E}(t) \cdot \mathbf{d}
\end{equation}
where $\mathbf{E}(t)$ is the electric field of a laser, and $\mathbf{d} = e\mathbf{r}$ is the electric-dipole operator. 

To find the matrix elements of this interaction in our basis, it is convenient to work in Cartesian coordinates. Choosing $z$ as the quantization axis, we first find $d_z$, the $z$-component of $\mathbf{d}$, which describes $\Delta m_F = 0$ or $\pi$ transitions.  
\begin{equation}
\begin{aligned}
\langle nJF M | d_z | n'J'F'M' \rangle =
\langle nJ || e {\bf r} || n'J'\rangle \\ 
\times \; (-1)^{M + J + I} \sqrt{(2 F + 1)(2F'+1) } \\
\times \left( 
\begin{array}{ccc}
F' & 1 & F \\
M' & 0 & -M
\end{array}
\right) 
\left\{
\begin{array}{ccc}
J & J' & 1 \\
F' & F & I
\end{array}
\right\} ,
\label{eq:dipole_operator_z}
\end{aligned}
\end{equation}
where the $( ::: )$ and $\{ ::: \}$ are the Wigner 3-j and 6-j symbols respectively, and the reduced matrix elements $\langle nJ || e {\bf r} || n'J'\rangle$ are known from the literature \footnote{We obtained all the matrix elements except three from \cite{Safronova2011}, elements between the $4d$ and $4f$ states were obtained directly in a private communication from M. S. Safronova of U. Delaware. All the elements we used are available online \cite{GitHubRef}.}. The $d_z$ matrix can be rotated to find the  $d_x$ and $d_y$ matrices:
\begin{equation}
\begin{aligned}
d_x = e^{i F_y \pi /2} d_z e^{- i F_y \pi /2} \\
d_y = e^{- i F_x \pi /2} d_z e^{i F_x \pi /2}
\label{eq:dipole_operator_x_y}
\end{aligned}
\end{equation}
where $F_x = (F_+ + F_-)/2 $ and $F_y=-i(F_+ - F_-)/2$ are total angular momentum components, given in terms of 
the ladder operators $F_\pm$ with matrix elements \cite{Sakurai}
\begin{equation}
\begin{aligned}
\langle nJF{M} | F_\pm | n'J'F'{M}' \rangle  = &
\sqrt{(F \mp M + 1) (F \pm M)} \\  
& \times \; \delta_{nJF,n'J'F'} 
\delta_{M,{M}' \pm 1}.
\end{aligned}
\end{equation}

The electric field is similarly described in Cartesian coordinates. As examples, if the incident optical field is monochromatic and polarized along $\mathbf{\hat{z}}$, the electric field is 
\begin{equation}
\mathbf{E_\pi}(t) =  \E    \mathrm{cos}(\omega t)  \mathbf{\hat{z}}
\end{equation}
where $\E$ is the amplitude of the electric field, $\omega = 2 \pi c / \lambda$ is the optical frequency, $c$ is the speed of light, and $\lambda$ is the wavelength.  Circularly-polarized light has the field 
\begin{equation}
\mathbf{E}_{\sigma^\pm}(t) = \frac{\E}{\sqrt{2}} \left[\mathrm{cos}(\omega t)\mathbf{\hat{x}}  \pm \mathrm{sin}(\omega t)\mathbf{\hat{y}} \right].
\end{equation}
The electric field of two linearly polarized fields with amplitudes ${\cal E}_i$, polarizations ${\bf n}_i$  frequencies $\omega_i$, $i\in\{1,2\}$, can be written
\begin{equation}
\mathbf{E}(t) =  \E_1 \mathrm{cos}(\omega_1 t)\mathbf{\hat{n}}_1 +  \E_2 \mathrm{cos}(\omega_2 t) \mathbf{\hat{n}}_2.
\end{equation}
It is important to note that the period $T$ in Eq. \ref{eq:U} refers to one period of the \textit{total} electric field, so we can calculate the light shifts due to multiple wavelengths as long as they are related by rational fractions. E.g. if $\lambda_1/\lambda_2 = a/b$, where $a$ and $b$ are positive integers, the period of the total electric field is the lowest common multiple of $T_1$ and $T_2$, where $T_i = 2 \pi / \omega_i = \lambda_i / c$ is the optical period, and $c$ is the speed of light.

In the above formulation  $H_0$ can be readily extended to include static magnetic and/or electric fields, and $V$ can be adapted to include magnetic and higher electric multipole transitions, provided the matrix elements are known. We acknowledge that here we neglect any possible vacuum field, relaxation,  continuum, or relativistic effects.

\begin{figure}
\centering
\includegraphics[width=0.9\columnwidth]{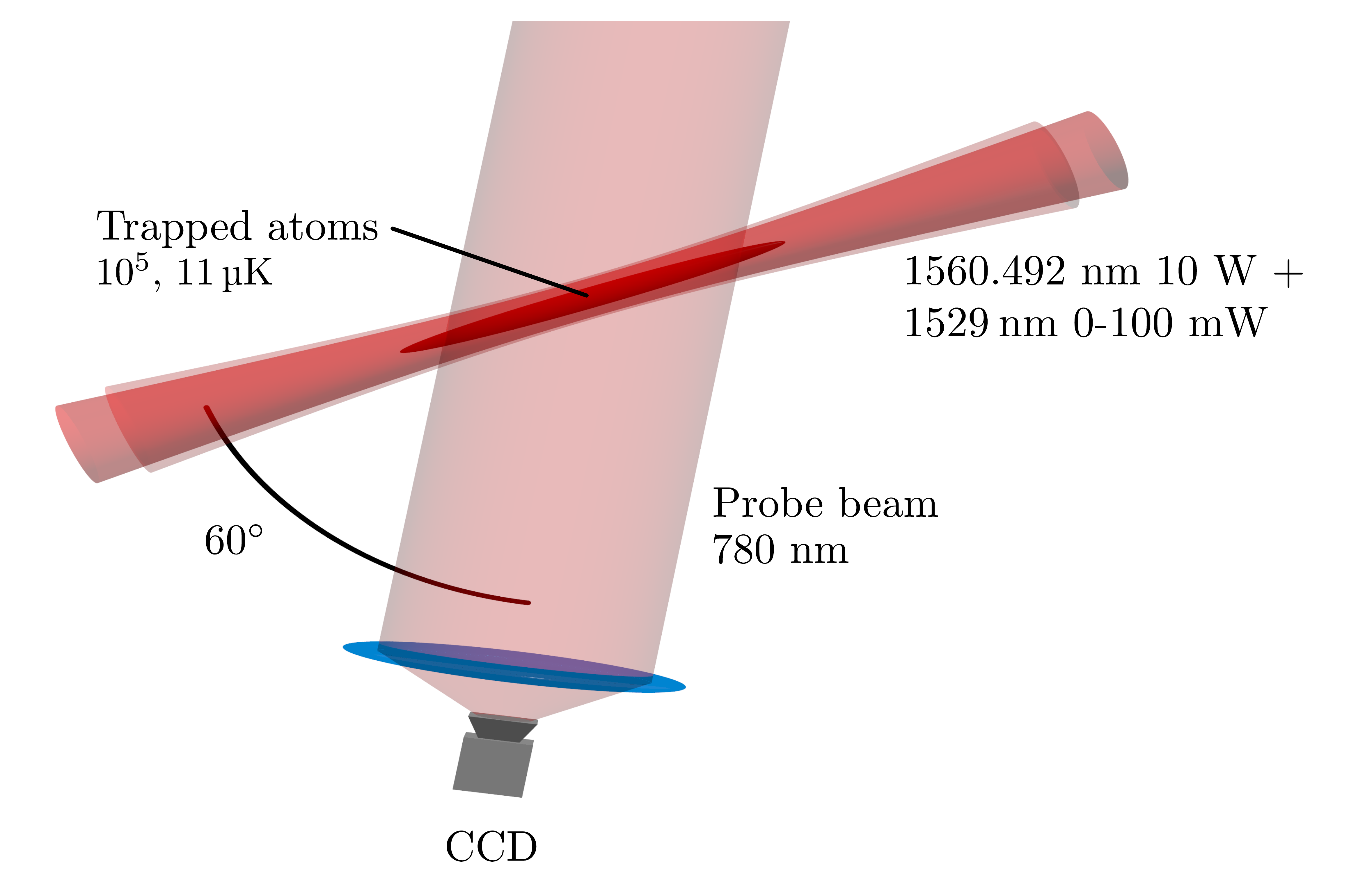}
\caption{Beam and atom cloud geometry in the experiment. A constant-power optical dipole trap at $\SI{1560.492}{\nano\meter}$ confines a cloud of $10^5$ atoms while a mode-matched beam around $\SI{1529}{\nano\meter}$ with adjustable power induces strong light shifts in the atoms. A probe beam with adjustable frequency at 780 nm is used for measuring absorption of the cloud as a function of frequency.} 
\label{fig:setup}
\end{figure}

\section{Experiment}
\label{section:experiment}
To validate the above numerical technique we perform spectroscopy of the D$_2$ hyperfine transitions in a cloud of cold $^{87}$Rb atoms in the presence of strong non-linear light shifts. A schematic of the experiment is shown in Fig. \ref{fig:setup}. To trap the atoms we use an optical dipole trap consisting of a single linearly polarized 10 W beam locked with $<$ 1 MHz stability to 1560.492 nm (the second harmonic of which is locked to a transition of the $^{87}$Rb D$_2$ line at 780.246 nm), and focused to a spot size of $\sim \SI{44}{\micro\meter}$. A second beam near 1529 nm is mode-matched to the 1560 nm beam, with a controllable power from 0--100 mW. The two beams are combined on a polarizing beamsplitter, with the 1560 reflected and the 1529 transmitted, so the polarizations are linear vertical and horizontal, respectively. The 1560 beam is not perfectly linear before the beamsplitter, and the polarization is not perfectly cleaned on reflection from the cube, so there is some residual ellipticity. The 1529 nm beam can be scanned across the $5P_{3/2} \rightarrow 4D_{3/2(5/2)}$ excited-state resonances at 1529.26 (1529.36) nm, so we can induce strong light shifts in the $5P_{3/2}$ states with relatively low intensities. We measured two datasets, one with the 1529 laser at 1529.282 nm and another at 1529.269 nm. The 1529 laser was not frequency-stabilised, and the wavelength was measured with a calibrated wavemeter to drift by $\pm 0.001 $ nm from the nominal wavelength over the duration of the measurements. A probe beam at 780 nm propagates at an angle of 60$^\circ$ relative to the trap axis, to reduce the chance of producing states that are ``dark'' to the probe light. The probe laser is stable to less than $\SI{100}{\kilo\hertz}$, and can be scanned up to $\SI{1}{\giga\hertz}$ to the red side of the D$_2$ transition.

\begin{figure}[t]
\centering
\includegraphics[width=0.8 \columnwidth]{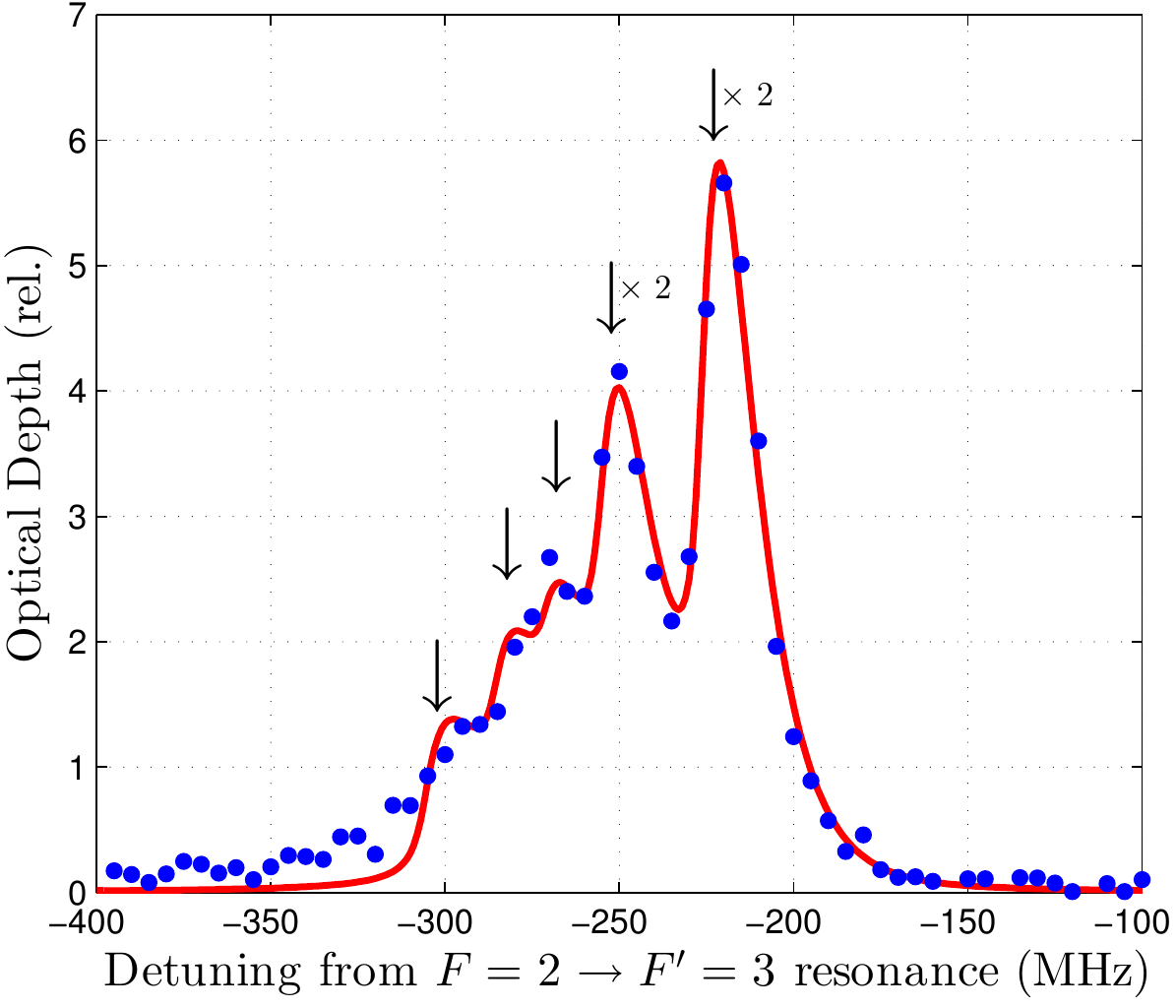}
\caption{Relative optical depth of atoms in the dipole trap around the light-shifted $F=2 \rightarrow F' = 3$ transition with only 1560 nm light. Blue dots show measured optical depth in a small transverse slice of the dipole trap, extracted from absorption images. Each point is from a single experiment run. The red line is a fit using Eq.~(\ref{eq:in_situ_abs}). The x-axis is relative to the free-space $|5S_{1/2}, F = 2 \rangle \rightarrow |5P_{3/2}, F = 3 \rangle$ transition. Free parameters in the fit were peak amplitudes, trap depth, and the ellipticity of trap light. Arrows show positions of resonances at maximum trap depth, ``$\times 2$" indicates 2 resonances within the width of the arrow.} 
\label{fig:1560_only}
\end{figure}

The experimental sequence is as follows: We trap approximately 3 $\times 10^6$ atoms in the $F=1$ ground state in the 1560 nm optical dipole trap. Initially the trap depth is about $\SI{270}{\micro\kelvin}$ and the atoms have a temperature of about $\SI{40}{\micro\kelvin}$. To ensure the atoms experience as homogeneous a light intensity as possible, we reduce the temperature and therefore the spatial extent of the cloud by performing an evaporation sequence followed by adiabatic increase of the trap depth back up to about $\SI{270}{\micro\kelvin}$, obtaining $10^5$ atoms at $\SI{11}{\micro\kelvin}$. We then pump the atoms into the $F = 2$ ground state and measure absorption of a probe laser as a function of the frequency of the probe beam and intensity of the 1529 nm beam. Fig. \ref{fig:1560_only} shows relative optical depth as function of probe beam frequency at zero 1529 nm beam intensity. We say ``relative'' as our image processing was calibrated for measuring the density of atoms in free space, correcting for saturation as described in \cite{Reinaudi2007}. Fig. \ref{fig:1560_1529a} shows relative optical depth as a function of both probe beam frequency and 1529.282 nm beam intensity, and Fig. \ref{fig:1560_1529b} shows the same but with 1529.269 nm light. The image processing technique described in \cite{Ockeloen2010} was found to help in detecting weak absorption signals.

\section{Results and Discussion}
\begin{figure}[t]
\centering
\includegraphics[width=0.8 \columnwidth]{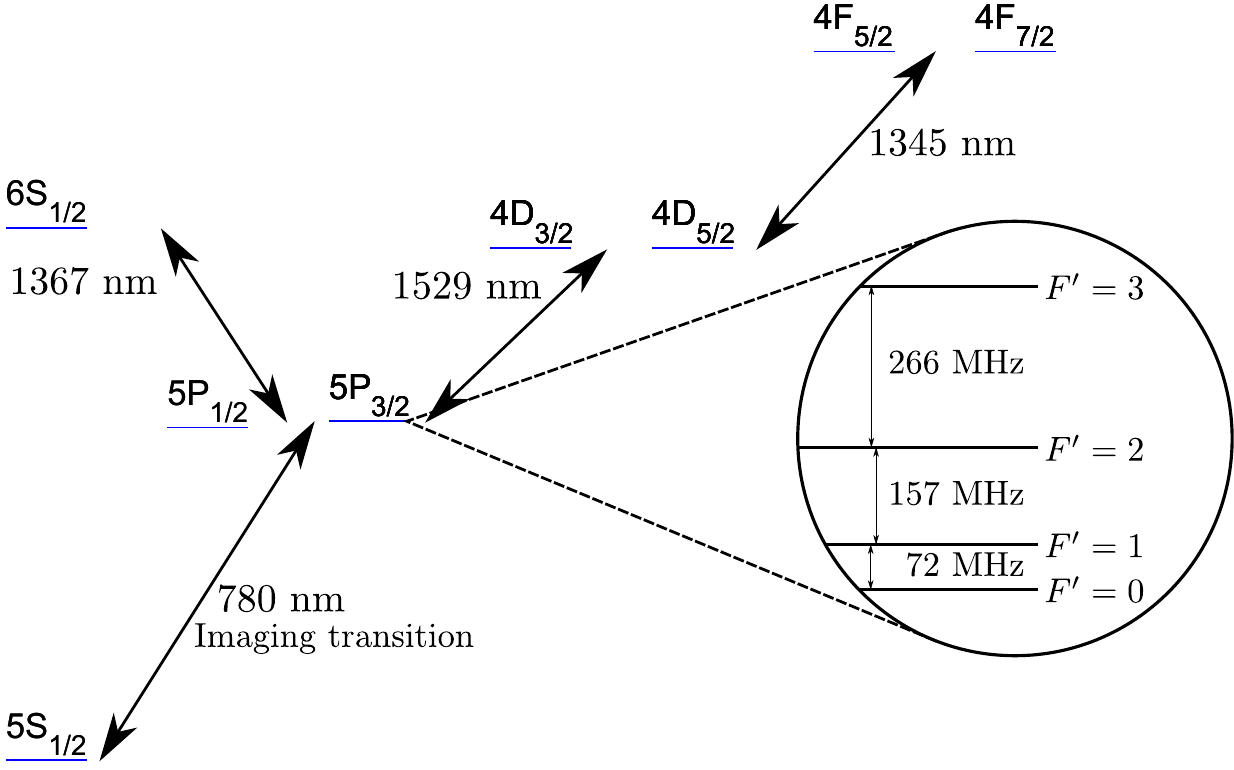}
\caption{$^{87}$Rb energy levels used for calculations in this paper. We performed several representative calculations with many more matrix elements up to $n = 10$ and found that these made a $< 1$ MHz contribution to the calculated light shifts under our experimental conditions, which is less than the uncertainty in our measurement. The inset shows the hyperfine splitting of the $|5P_{3/2}\rangle$ levels. We included hyperfine splitting for all levels except the $4f$ levels, for which we were unable to find hyperfine constants in the literature. } 
\label{fig:energy_levels_labelled}
\end{figure}

\subsection{Light Shifts @ 1560 nm}
We first consider in detail the absorption spectrum of atoms in the trap with no other incident light, i.e. light shifts induced just by the $\SI{1560.492}{\nano\meter}$ trap itself (Fig. \ref{fig:1560_only}). We adapt an equation from \cite{Brantut2008} as a model for our signal. Eq.~(\ref{eq:in_situ_abs}) describes theoretical optical depth as a function of probe detuning of a mixture of populations of non-interacting two-level atoms with differential light shift at thermal equilibrium in a harmonic potential. The model treats each possible transition as a separate population of two-level atoms, so the transition frequency of each population corresponds to a transition from the five near-degenerate $|5S_{1/2}, F = 2\rangle$ ground states to each of the seven light-shifted $|5P_{3/2}, F = 3\rangle$  excited states.

\begin{equation}
A(\delta) = \sum_{i = 1}^{7} C_i \int_0^{\infty} \frac{u^2 e^{-u^2} du}{1 + 4 ( \delta + \nu_i - t_i u^2)^2} 
\label{eq:in_situ_abs}
\end{equation}
where $i$ indicates the $i^{th}$ state, $t_i = \frac{k_B T}{\hbar \Gamma}\left( \frac{\alpha_{e,i}}{\alpha_g} - 1 \right)$, $\nu_i = \frac{U}{\hbar \Gamma} \frac{\alpha_{e,i}}{\alpha_g} $, and $\delta = \frac{\omega - \omega_0}{\Gamma}$ are normalised temperature, trap depth, and probe detuning, respectively. $k_B$ is Boltzmann's constant, $T$ is the cloud temperature ($\SI{11}{\micro\kelvin}$, from time-of-flight measurements), $\Gamma$ is the natural linewidth, $\alpha_{e,i}$ is the polarizability of the $i^{th}$ excited state, $\alpha_g$ the ground state (the tensor light shift of the different $|5S_{1/2}, F = 2\rangle$ ground states is on the order of kHz), $\omega$ is the probe laser frequency, $\omega_0$ is the free-space transition frequency, and $m$ is the mass of the $^{87}$Rb atom. The differential polarizability of a transition is equal to the differential light shift, i.e. $\alpha_e/\alpha_g = \Delta f_e/\Delta f_g$, where $\Delta f_{e(g)}$ is the light shift of the excited (ground) state, $U$ is the trap depth, and $C_i$ is a fitting parameter depending on the number of atoms measured, and the absorption cross-section of the $i^{th}$ level for the probe beam. 

To fit Eq.~(\ref{eq:in_situ_abs}) to the data shown in Fig. \ref{fig:1560_only} we model the electric field of the 1560 laser as
\begin{equation}
\mathbf{E}_1(t) = \frac{\E_1}{\sqrt{2}}(\mathrm{cos}(\omega_1 t) \mathbf{\hat{x}} + \mathrm{cos}(\omega_1 t + \phi) \mathbf{\hat{y}}),
\end{equation}
and calculate the light shifts as described in section \ref{section:theory}, to obtain the differential light shift and consequently the differential polarizability $\alpha_e/\alpha_g$. We include the quadrature phase $\phi$ to account for a slight ellipticity of the 1560 light after reflection at a polarizing beamsplitter as discussed in section \ref{section:experiment}. If $\phi = 0$ this simply describes a linearly polarized electric field oscillating in the $\mathbf{\hat{x}} + \mathbf{\hat{y}}$ plane. The coefficients $C_i$, electric field $E_1$, and quadrature phase $\phi$ were free parameters in the fit \footnote{Quantitative prediction of $C_i$ is feasible but would require use of the optical Bloch equations to solve for atomic dynamics in the presence of the 1560 beam, the single probe beam at 780 nm, and repump light also at 780 nm which is emitted from six directions toward the centre of the trap.}. The light intensity is related to the electric field by
\begin{equation}
I = \frac{\epsilon_0 c}{2} |E|^2
\end{equation}
where $\epsilon_0$ is the permittivity of free space and $c$ is the speed of light. From the fit we extracted $I_{1560} = \SI{2.91+-0.01e9}{\watt\meter^{-2}}$, which agrees well with power meter measurements, and $\phi = 0.133 \pm 0.009$. By using colder atoms and/or a deeper trap, these quantities could be known more accurately. The trap depth $U$ is equal to the light shift of the ground state at peak light intensity at the center of the trap. We obtained $U = h \cdot 5.623 \pm \SI{0.004}{\mega\hertz}$  ($= k_B \cdot 270.0 \pm \SI{0.2}{\micro\kelvin})$.  We can compare the $U$ obtained from the fit to $U_{calc}$ calculated from the measured trap oscillation frequency $f_{osc} = \SI{1.22}{\kilo\hertz}$ and the beam waist measured with a beam profiler $w = \SI{44}{\micro\meter}$ as $U_{calc} = (2 \pi w f_{osc})^2 m/4 = h \cdot \SI{6.2}{\mega\hertz}$. The difference between the two can be explained with an error in the measurement of the beam waist of $\SI{2}{\micro\meter}$.

The arrows in Fig. \ref{fig:1560_only} show calculated light shifts of atomic transitions at the bottom of the trap, i.e. $\Delta f_{e,i} - \Delta f_g$ at peak light intensity. The data peaks are slightly offset from the theoretical peaks due to the finite temperature of the atoms: atomic density peaks above the bottom of the trap. 

For our calculations we used only the energy levels shown in Fig. \ref{fig:energy_levels_labelled}, comprising 136 distinct states. We performed several representative calculations with levels up to $n = 10$ and found these extra states contributed less than 1 MHz to the calculated light shifts. We computed $\mathscr{U}$ numerically with Eq.~(\ref{eq:U}) and cut off $N$ at some finite value,  but making sure it is sufficiently high such that the result has converged. For calculations with the 1560 beam only we used $N = 200$. All calculations were done in MATLAB and our code is available online \cite{GitHubRef}.

\begin{figure}
\centering
\includegraphics[width=0.8 \columnwidth]{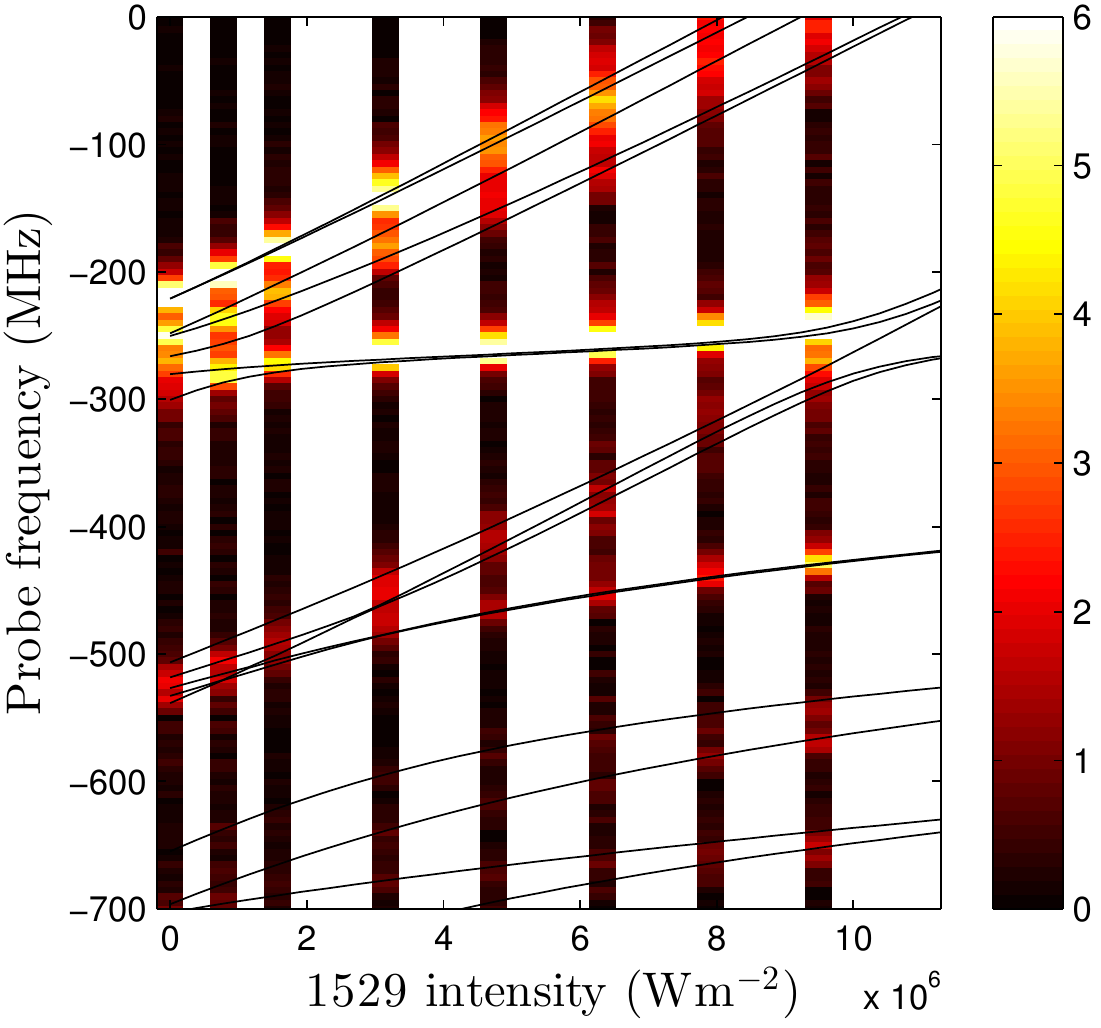}
\caption{Relative optical depth of the cloud with \SI{1560.492}{\nano\meter} and \SI{1529.282}{\nano\meter} incident light. The probe frequency is relative to the free-space $|5S_{1/2}, F=2 \rangle \rightarrow |5P_{3/2}, F=3 \rangle$ transition. The black lines show calculated energy levels of dressed states. Shading shows measured optical depth of the atomic cloud in arbitrary units with the scale shown in the colour bar on the right. Each column is scaled to have the same maximum value. After using the data shown in Fig. \ref{fig:1560_only} as a calibration of the experimental parameters, the only fitting parameter here is the calibration of the 1529 nm beam power.} 
\label{fig:1560_1529a}
\end{figure}

\begin{figure}
\centering
\includegraphics[width=0.8 \columnwidth]{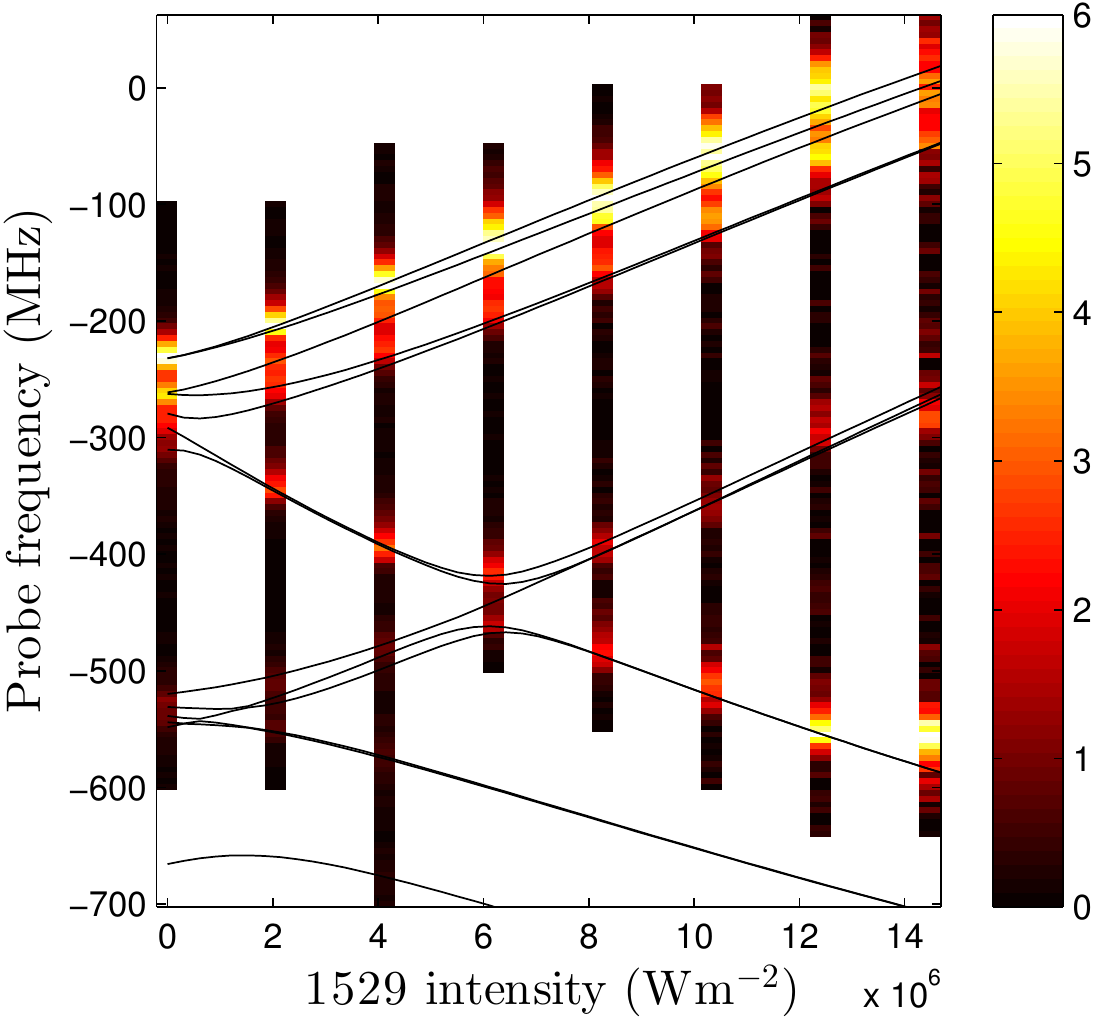}
\caption{Relative optical depth with \SI{1560.492}{\nano\meter} and \SI{1529.269}{\nano\meter} incident light, representation as in Fig. \ref{fig:1560_1529a}.  Here, because the  \SI{1529}{\nano\meter} wavelength is closer to resonance the nonlinearity is clearer, including avoided and non-avoided crossings. Because the ratio of wavelengths does not form a simple rational fraction, black theory curves are calculated for the wavelength $\lambda_1 =  \SI{1559.854}{\nano\meter}$ (i.e. $\lambda_1 = \frac{51}{50} \lambda_2$). We mostly compensate for this mismatch between the real and assumed wavelength of the 1560 nm light by reducing the intensity of light in the calculation by 2.6\%. There is still an estimated error of up to 150 kHz in the calculated light shifts, however this is below the resolution of our measurement. } 
\label{fig:1560_1529b}
\end{figure}

\subsection{Light Shifts @ 1560 nm + 1529 nm}
Figs. \ref{fig:1560_1529a} and \ref{fig:1560_1529b} show absorption of the probe beam as a function of probe frequency and 1529 beam intensity. The black lines are calculated transition frequencies relative to the free-space $|5S_{1/2}, F = 2 \rangle \rightarrow |5P_{3/2}, F = 3 \rangle$ transition. We used the measurement shown in Fig. \ref{fig:1560_only} as a calibration of the experimental parameters to then perform the calculation of energy level shifts as a function of 1529 beam intensity, so the only fitting parameter here is the 1529 intensity. The left-most column is the same data as that shown in Fig. \ref{fig:1560_only}.

For calculating light shifts with both the 1560 nm and 1529 nm beams present we model the electric field as
\begin{equation}
\mathbf{E}(t) = \mathbf{E}_1(t) + \frac{\E_2}{\sqrt{2}}[\mathrm{cos}(\omega_2 t)\mathbf{\hat{x}}  - \mathrm{cos}(\omega_2  t)\mathbf{\hat{y}}]
\end{equation}
which describes the electric field of the 1560 beam added to the linearly polarized 1529 beam. The two fields are orthogonally polarized if $\phi = 0$. The wavelength of the 1560 trapping beam was $ \lambda_1 = \SI{1560.492}{\nano\meter}$, so for one measurement we set the wavelength of the 1529 beam to be $\lambda_2 = \frac{49}{50}\lambda_1 = \SI{1529.282}{\nano\meter}$. For another measurement we set $\lambda_2 = \SI{1529.269}{\nano\meter}$, and modelled the 1560 wavelength as $\lambda_1 = \frac{51}{50}\lambda_2$. The 1529 nm beam was measured to drift by $\pm \SI{0.001}{\nano\meter}$ over the duration of the measurements, which can explain the deviation of the data from the theory. 

\section{Measuring electric-dipole matrix elements}
A potential application of these techniques is precision measurement of excited-state electric-dipole matrix elements. The standard technique of measuring lifetimes is complicated for excited-state transitions by the presence of multiple decay channels, although techniques exist such as measuring the relative light shifts at two different wavelengths \cite{Sahoo2009} (the technique presented in that work requires atoms with  conveniently-placed metastable states), or using strong magnetic fields to ``isolate" simpler level structures \cite{Whiting2016}. We propose a complementary method to measure the dipole matrix element of any excited-state transition by using one or more lasers to couple the higher level of an imaging transition to an excited-state transition. Light near-resonant with the excited-state transition induces strong light shifts, affecting the frequency of the imaging transition, which can be measured and compared to theory. 

To test this idea, we calculated light shifts under a representative set of our experimental conditions, and then adjusted the value of each electric-dipole matrix element used in our calculation by 0.1\%. We found the light shifts depend strongly on the values of the  $\langle 5P_{3/2} || e {\bf r} || 4 D _{3/2}\rangle$ and $\langle 5P_{3/2} || e {\bf r} || 4 D _{5/2}\rangle$ elements, but are hardly at all dependent on the value of any other matrix element. A change of either of these matrix elements by 0.1\% changes the calculated light shifts of all the $|5P_{3/2}\rangle$ hyperfine levels by at least 100 kHz, suggesting that measuring to this accuracy would constrain these matrix elements to 0.1\%, better than their current known precision.

One could then add additional light at 1345 nm (see Fig. \ref{fig:energy_levels_labelled}), which would make the shift of the $|5P_{3/2}\rangle$ levels now also dependent on the $\langle 4D || e {\bf r} || 4F \rangle$ matrix elements, as these levels are now coupled by a ``ladder" of near-resonant light. In this way measuring dipole matrix elements reduces to measuring energy levels and comparing to theory.

Measuring stronger light shifts would enable more accurate determination of the relevant matrix element, and using multiple wavelengths enables measurement of excited-state transitions which are normally difficult to access. The matrix element to measure can be selected by choice of wavelength(s). 

One potential difficulty lies in determining the in-situ light intensity, which is difficult to measure independently. The light intensity could be included as a free parameter in fitting theory to data, but the Hamiltonian, and therefore the light shifts, depend only on the product of the electric field and the electric-dipole transition matrix, i.e. $V(t) = \mathbf{E}(t) \cdot \mathbf{d}$. This means that if all the electric dipole matrix and light intensity are free parameters in fitting theory to data, the only constraint is the product of the two terms, not the absolute value of either. However ensuring that the light shifts depend also on a known dipole matrix element that is not a free parameter in the fit would remove this ambiguity. 

\section{Conclusion}
We have presented a theory for the calculation of strong atomic light shifts due to multiple wavelengths, where the shifts can be nonlinear and larger than the hyperfine splitting. We validated our theory by predicting and measuring light shifts of the D$_2$ transition in $^{87}$Rb caused by incident light nearly resonant with the $5P_{3/2} \rightarrow 4D$ transitions.

\begin{acknowledgments}
Work supported by MINECO/FEDER, MINECO projects MAQRO (Ref. FIS2015-68039-P), XPLICA (FIS2014-62181-EXP) and Severo Ochoa grant SEV-2015-0522, Catalan 2014-SGR-1295, by the European Union Project QUIC (grant agreement 641122), European Research Council project AQUMET (grant agreement 280169) and ERIDIAN (grant agreement 713682), and by Fundaci\'{o} Privada CELLEX. The authors would also like to thank J. Douglas and R. Jim\'enez-Mart\'inez for useful feedback on the manuscript.
\end{acknowledgments}

\bibliography{stark_shifts_article}

\end{document}